\documentclass{article}
\newcommand{\be}{\begin{equation}}
\newcommand{\ee}{\end{equation}}
\newcommand{\nbar}[1]{\overline{#1}}                       

\def\bea{\begin{eqnarray}}
\def\eea{\end{eqnarray}}
\def\beas{\begin{eqnarray*}}
\def\eeas{\end{eqnarray*}}
\def\sla{\raise.15ex\hbox{$/$}\kern-.57em}

\def\parp{\partial^+}

\usepackage{amsmath}
\usepackage{latexsym}

\begin{document}
\begin{titlepage}
\begin{flushright}    UFIFT-HEP-05-09 \\ 
\end{flushright}
\vskip 1cm
\centerline{\LARGE{\bf { Non-linear Realization  of }}}
\vskip .5cm
\centerline{\LARGE{\bf {$PSU(2,2\,\vert\,4)$ on the Light-Cone  }}}

\vskip 1.5cm
\centerline{\bf Sudarshan Ananth,} 
\vskip .2cm
\centerline{\em  Institute for Fundamental Theory,}
\centerline{\em Department of Physics, University of Florida}
\centerline{\em Gainesville FL 32611, USA}
\vskip .5cm
\centerline{\bf Lars Brink,  }
\vskip .2cm
\centerline{\em Department of Fundamental Physics}
\centerline{\em Chalmers University
of Technology, }
\centerline{\em S-412 96 G\"oteborg, Sweden}

\vskip .5cm
\centerline{\bf Sung-Soo Kim and Pierre Ramond  }
\vskip .2cm
\centerline{\em  Institute for Fundamental Theory,}
\centerline{\em Department of Physics, University of Florida}
\centerline{\em Gainesville FL 32611, USA}

\vskip 1.5cm

\centerline{\bf {Abstract}}
\vskip .5cm
\noindent  The symmetries of the ${\mathcal N}=4$ SuperYang-Mills theory on the light-cone are discussed, solely in terms of its physical degrees of freedom. We derive explicit expressions for the generators of the  $PSU(2,2\,\vert\,4)$ superalgebra, both in the free theory, and to all orders in the gauge coupling of the classical theory. We use   these symmetries to construct  its Hamiltonian, and show that it can be written as a quadratic form of a fermionic superfield. 
\vfill
\begin{flushleft}
May 2005 \\
\end{flushleft}
\end{titlepage}

\section{Introduction}
No-go theorems show that massless fields of helicity higher than two cannot interact {\em locally} with gravity, and respect Lorentz invariance~\cite{WW}.  However on the light-cone, Bengtsson, Bengtsson and Brink~\cite{BBB}  used algebraic consistency to build  Lorentz-invariant cubic interactions of massless particles with arbitrary helicities.  Although algebraically demanding, this light-cone approach seems natural to construct Lorentz invariant interactions, as infinite towers of massless higher-spin fields~\cite{dreams} have emerged in possible generalizations of ${\cal N}=1$ Supergravity in eleven dimensions.  

Supergravity in eleven dimensions~\cite{Julia}, when reduced to four space-time dimensions, becomes the maximally supersymmetric ${\cal N}=8$ Supergravity. Its light-cone formulation in superspace  is not entirely known, as its four- and higher-point interactions have not yet been written in terms of one light-cone superfield.  

A purpose of this paper is to devise algebraic techniques for finding these higher point functions. This is quite complicated, but fortunately, there is a much simpler maximally supersymmetric, but challenging theory, in which one can study these questions:  ${\cal N}=4$ SuperYang-Mills in four dimensions. It is very similar in structure to  ${\cal N}=8$ supergravity. Both are elegantly described on the light-cone by one superfield containing only physical degrees of freedom, and  both oxidize~\cite{ABR1,ABR2} naturally to their higher-dimensional progenitors:  ${\cal N}=4$ to its ten-dimensional ${\cal N}=1$ parent, and  ${\cal N}=8$ to ${\cal N}=1$ supergravity in eleven dimensions, producing superspace descriptions without auxiliary fields (although this has been shown only to first order in $\kappa$ for supergravity).

The ${\cal N}=4$ Yang-Mills theory, where the four-point function is well-known, is a natural testing ground for developing algebraic tools to  generate  higher-point interactions. We expect that the techniques we have uncovered will enable us to determine the full ${\cal N}=8$ supergavity Lagrangian.

We first review the light-cone description of the  ${\cal N}=4$ theory, and display the classical symmetries of its {\em free} action, which act  {\em linearly} on  the superfield. These include not only the superPoincar\'e transformations, but also the superconformal symmetries, which combine with an internal $SU(4)$ to form the $PSU(2,2\,\vert\,4)$ superalgebra.

The action for ${\cal N}=4$ Yang-Mills is well known, written  in terms of a single chiral superfield that encapsulates  its physical degrees of freedom: one helicity-one gauge field, four helicity one-half  fermions and their conjugates, and six helicity zero scalar fields. The same light-cone formulation was used to prove its ultraviolet finiteness~\cite{BLN}. 

Space-time symmetries  split into two types; kinematical symmetries which are not altered by interactions, and dynamical symmetries which are realized {\em non-linearly} on the fields. In a Lorentz invariant field theory, these dynamical symmetries are the light-cone time translation generated by the Hamiltonian, and the boosts. In supersymmetric theories, the supersymmetries also split,  with the dynamical supersymmetries  acting as  the ``square-roots" of the Hamiltonian. The  same split occurs in the superconformal transformations.   

The construction of these non-linear transformations was initiated in reference [2],  order by order in the coupling constant. When a gauge theory is expressed solely in terms of its physical degrees of freedom, Poincar\'e invariance is not manifest, and remains to be checked. Bengtsson {\em et al} formed  Ans\"atze for both  boosts and Hamiltonian  variations, and verified  closure of the Super-Poincar\'e algebra to first order in the coupling constant $g$. They reproduced the well-known cubic interaction of the ${\cal N}=4$ theory, and showed how closure requires several superfields linked by an antisymmetric  $f^{abc}$, where $a,\,b,\, c$ label the superfields.  

We complete their program to order $g^2$. We first show that the algebraic constraints fix the form of the dynamical supersymmetry transformations {\em uniquely}, with no order $g^2$ corrections. From the fact that the conformal group is simple,  its kinematical symmetries, together with the form of the dynamical supersymmetry transformations,  fully determine the entire $PSU(2,2\,\vert\,4)$  with all classical interactions included! We show how  these transformations  generate the complete classical Hamiltonian, including the four-point interaction. The full antisymmetry of the structure functions and their Jacobi identities is required by the algebra. In this light-cone language, implementation of the space-time symmetries requires gauge symmetries, and naturally reconstructs the required Lie algebra structures.

Finally, we show that  the Hamiltonian of ${\cal N}=4$ Yang-Mills can be written as a {\em quadratic form}  of a {\em fermionic} superfield. This fermionic superfield is simply the dynamical supersymmetry variation of the original superfield. 
 
In  future publications, we hope to extend these techniques to derive the form of the quartic and higher-point interactions in ${\cal N}=8$ supergravity. 
 
\renewcommand{\theequation}{2.\arabic{equation}}
\setcounter{equation}{0}
\section{Light-Cone Formulation: Review}

\subsection{Notation}
With the space-time metric $(-,+,+,\dots,+)$, the light-cone coordinates  and their derivatives are 

\bea
{x^{\pm}}&=&\frac{1}{\sqrt 2}\,(\,{x^0}\,{\pm}\,{x^3}\,)\ ;\qquad ~ {\partial^{\pm}}=\frac{1}{\sqrt 2}\,(\,-\,{\partial_0}\,{\pm}\,{\partial_3}\,)\ ; \\
x &=&\frac{1}{\sqrt 2}\,(\,{x_1}\,+\,i\,{x_2}\,)\ ;\qquad  {\bar\partial} =\frac{1}{\sqrt 2}\,(\,{\partial_1}\,-\,i\,{\partial_2}\,)\ ; \\
{\bar x}& =&\frac{1}{\sqrt 2}\,(\,{x_1}\,-\,i\,{x_2}\,)\ ;\qquad  {\partial} =\frac{1}{\sqrt 2}\,(\,{\partial_1}\,+\,i\,{\partial_2}\,)\ ,
\eea
so that 

\be
{\parp}\,{x^-}={\partial^-}\,{x^+}\,=\,-\,1\ ;\qquad {\bar \partial}\,x\,=\,{\partial}\,{\bar x}\,=+1 \ .
\ee
In four dimensions, massless particles with helicity can be described by a  complex  field, and its complex conjugate of opposite helicity. Particles with no helicity are described by real fields. 

The particle content of the ${\cal N}=4$ Yang-Mills theory is best described in ten dimensions, where the ${\cal N}=1$ supermultiplet  contains  eight vectors 
 and  eight spinors of the little group  $SO(8)$. Under the decomposition 

\be
SO(8)\supset~SO(2)~\times~SO(6)\ ,
\ee
these give 

\be
{\bf 8}^{}_v~=~{\bf 6}^{}_0+{\bf 1}^{}_1+{\bf 1}^{}_{-1}\ ,\qquad 
{\bf 8}^{}_s~=~{\bf 4}^{}_{1/2}+{\bf \bar4}^{}_{-1/2}\ ,
\ee
where the  subscripts denote the $SO(2)$ helicity, showing that  the $\mathcal N=4$ theory contains  six scalar fields, a  vector field and four spinor fields and  their conjugates. To describe them in compact notation, we  introduce anticommuting Grassmann variables $\theta^m$ and $\bar\theta_m$, 

\be
\{\,{\theta^m}\,,\,{{ \theta}^n}\,\}\,~=~\,\{{{\bar \theta}_m}\,,\,{\bar\theta_n}\,\}\,~=~\,\{{{\bar \theta}_m}\,,\,{\theta^n}\,\}\,~=~0\ ,
\ee
which transform as the spinor representations of $SO(6)\sim SU(4)$,

\be
\theta^m_{}~\sim~ {\bf 4}^{}_{1/2}\ ;\qquad \overline\theta_{}^m~\sim~ {\bf \bar 4}^{}_{-1/2}\ ,
\ee
where $m,n, \dots  =1\,,\dots\, 4$. Their derivatives are written as 

\be
{{\bar \partial}_m}\,~\equiv~\,\frac{\partial}{\partial\,{\theta^m}}\ ;\qquad{\partial^m}\,~\equiv~\,\frac{\partial}{\partial\,{\bar \theta}_m}\ ,
\ee
and satisfy

\be
\{\,{\partial^m}\,,\,{{\bar \theta}_n}\,\}\,~=~\,{{\delta^m}_n}\ ;\qquad \{\,{{\bar \partial}_m}\,,\,{\theta^n}\,\}\,~=~\,{{\delta_m}^n}\ .
\ee

\subsection{Superfield Action} 
{\it All} the physical degrees of freedom of the $\mathcal N=4$ theory can be captured in a single  complex superfield~\cite{BLN2} 

\bea
\phi\,(y)&=&\frac{1}{ \partial^+}\,A\,(y)\,+\,\frac{i}{\sqrt 2}\,{\theta_{}^m}\,{\theta_{}^n}\,{\nbar C^{}_{mn}}\,(y)\,+\,\frac{1}{12}\,{\theta_{}^m}\,{\theta_{}^n}\,{\theta_{}^p}\,{\theta_{}^q}\,{\epsilon_{mnpq}}\,{\partial^+}\,{\bar A}\,(y)\cr
& &~~~ +~\frac{i}{\partial^+}\,\theta^m_{}\,\bar\chi^{}_m(y)+\frac{\sqrt 2}{6}\theta^m_{}\,\theta^n_{}\,\theta^p_{}\,\epsilon^{}_{mnpq}\,\chi^q_{}(y) \ .
\eea
In this notation, the eight original gauge fields $A_i\ ,i=1,\dots,8$ appear as

\be
A~=~\frac{1}{\sqrt 2}\,(A^{}_1+i\,A^{}_2)\ ,\qquad \bar A~=~\frac{1}{\sqrt 2}\,(A^{}_1-i\,A^{}_2) \ ,
\ee
while the six scalar fields are written as antisymmetric $SU(4)$ bi-spinors 

\bea
C_{}^{m\,4}~=~\frac{1}{\sqrt 2}\,({A^{}_{m+3}}\,+\,i\,{A^{}_{m+6}})\ ,\qquad \nbar C^{}_{m\,4}~=~\frac{1}{\sqrt 2}\,({A^{}_{m+3}}\,-\,i\,{A^{}_{m+6}})\ ,
\eea
for $m\;\neq\,4$; complex conjugation is akin to duality, 

\bea
\label{dual}
{{\nbar C}^{}_{mn}}~=~\,\frac{1}{2}\,{\epsilon^{}_{mnpq}}\,{C_{}^{pq}} \ .
\eea
The fermion fields are denoted by $\chi^m$ and $\bar\chi_m$. All these fields carry adjoint indices of the gauge group (not shown here), and are local in the modified light-cone coordinates  

\bea
y~=~\,(\,x,\,{\bar x},\,{x^+},\,y^-_{}\equiv {x^-}-\,\frac{i}{\sqrt 2}\,{\theta_{}^m}\,{{\bar \theta}^{}_m}\,)\ .
\eea
In this particular light-cone formulation called $LC_2$,  all the unphysical degrees of freedom have been integrated out, leaving only  the physical ones. 

We introduce chiral derivatives,  

\bea
{d^{\,m}}=-{\partial^m}\,-\,\frac{i}{\sqrt 2}\,{\theta^m}\,{\partial^+}\ ;\qquad{{\bar d}_{\,n}}=\;\;\;{{\bar \partial}_n}\,+\,\frac{i}{\sqrt 2}\,{{\bar \theta}_n}\,{\partial^+}\ ,\eea
which satisfy the anticommutation relations

\be
\{\,{d^m}\,,\,{{\bar d}_n}\,\}\,=\,-i\,{\sqrt 2}\,{{\delta^m}_n}\,{\parp}\ .
\ee
One verifies that $\phi$ and its complex conjugate $\bar\phi$ satisfy the chiral constraints

\be
{d^{\,m}}\,\phi\,=\,0\ ;\qquad {\bar d_{\,m}}\,\bar\phi\,=\,0\ ,
\ee
as well as the ``inside-out" relations

\be
\bar d_m^{}\,\bar d_n^{}\,\phi~=~\frac{1}{ 2}\,\epsilon_{mnpq}^{}\,d^p_{}\,d^q_{}\,\bar\phi\ ,
\ee
\be
 d^m_{}\, d^n_{}\,\bar\phi~=~\frac{1}{ 2}\,\epsilon^{mnpq}_{}\,\bar d_p^{}\,\bar d_q^{}\,\phi\ .
\ee
The Yang-Mills action is then simply

\be
\int d^4x\int d^4\theta\,d^4 \bar \theta\,{\cal L}\ ,
\ee
where

\bea
{\cal L}&=&-\bar\phi\,\frac{\Box}{\partial^{+2}}\,\phi
~+\frac{4g}{3}\,f^{abc}_{}\,\Big(\frac{1}{\partial^+_{}}\,\bar\phi^a_{}\,\phi^b_{}\,\bar\partial\,\phi^c_{}+\frac{1}{\partial^+_{}}\,\phi^a_{}\,\bar\phi^b_{}\,\partial\,\bar\phi^c_{}\Big)\cr
&&-g^2f^{abc}_{}\,f^{ade}_{}\Big(\,\frac{1}{\partial^+_{}}(\phi^b\,\partial^+\phi^c)\frac{1}{\partial^+_{}}\,(\bar \phi^d_{}\,\partial^+_{}\,\bar\phi^e)+\frac{1}{2}\,\phi^b_{}\bar\phi^c\,\phi^d_{}\,\bar\phi^e\Big)\ .
\label{ACTION}\eea
Grassmann integration  is normalized so that $\int d^4\theta\,\theta^1\theta^2\theta^3\theta^4=1$, and $f^{abc}$ are the structure functions of the Lie algebra. It is straightforward to verify that this action reproduces the component action for the $\mathcal N=4$ theory, by simply performing the Grassmann integrals. In this form,  SuperPoincar\'e invariance is far from obvious,  the price for having eliminated the unphysical degrees of freedom. 

\renewcommand{\theequation}{3.\arabic{equation}}
\setcounter{equation}{0}
\section{Symmetries of $\mathcal N=4$ SuperYang-Mills}
The $\mathcal N=4$ theory is invariant under transformations generated by the superalgebra $PSU(2,2\,\vert\,4)$. Its $30$ bosonic generators describe the conformal group $SO(4,2)$ and an internal $SU(4)$;  its $32$ fermionic generators consist of four supersymmetries and superconformal symmetries and their conjugates. 

Some of the $PSU(2,2|4)$ generators have been discussed in papers
by Beltisky {\em et al} \cite{Belt}, who have used the above formalism
to study the integrability properties of the dilatation operator. In
this paper we construct the full superconformal algebra in its light-cone form.
Since the theory is ultraviolet finite, this symmetry survives quantization, and is therefore of great interest. The SuperPoincar\'e subalgebra  of $PSU(2,2\,|\,4)$ is well-known, its  generators having been constructed in references \cite{BBB,ABR1}. In the following section, we review their construction, and build the rest of the  superconformal algebra (on the light supercone). 


\subsection{SuperPoincar\'e Algebra}
The  light-cone Poincar\'e generators split into two types,  kinematical and dynamical. Dynamical generators  are those which contain the ``light-cone time derivative". Three  of the momenta are kinematical

\be
 p^+_{}~=~-i\,\partial^+_{}\ ,\qquad p~=~-i\,\partial\ ,\qquad \bar p~=~-i\,\bar\partial\ ,
\ee
while the fourth,  the light-cone Hamiltonian, shown here without interactions,

\be
p^-_{}~=~-i\frac{\partial\bar\partial}{\partial^+_{}}\ ,\ee
is dynamical. The Lorentz generators include the kinematical transverse space rotation
\be
j~=~x\,\bar\partial-\bar x\,\partial+\frac{1}{ 2}\,(\,{\theta^p}\,{{\bar \partial}_p}\,-\,{{\bar \theta}_p}\,{\partial^p}\,)\,-\lambda\ ,
\ee
where

\be
\lambda~=~\frac{i}{4\sqrt{2}\,\partial^+}\,(\,d^p\,\bar d_p-\bar d_p\,d^p\,)\ ,
\ee
 measures the helicity of the superfield ($\lambda\,=\,+\,1$ for a chiral superfield and $-\,1$ for an anti-chiral superfield). This form ensures that the chirality constraints are preserved. Under a transverse space rotation, this generator acts as a differential operator on the chiral superfield

\be
\delta\,\phi~=~i\,\omega\,j\,\phi\ ,\qquad \delta\,\bar \phi~=~-i\,\omega\,j\,\bar \phi\ .
\ee
 Acting on $\phi$, the remaining kinematical generators read 

\be
j^+_{}~=~i\, x\,\partial^+_{}\ ,\qquad \bar j^+_{}~=~i\,\bar x\,\partial^+_{}\ ,\qquad  j^{+-}_{}~=~i\,x^-_{}\,\partial^+_{}-\frac{i}{2}\,(\,\theta^p_{}\bar\partial^{}_p+\bar\theta^{}_p\,\partial^p_{}\,)+i\ .
\ee
These generators preserve chirality,   since 

\be
[\,j^{+-}_{}\,,\,y^-_{}\,]~=~-i\,y^-_{}\ ,\qquad [\,j^{+-}_{}\,,\,d^m_{}\,]~=~\frac{i}{2}\,d^m_{}\ ,\qquad [\,j^{+-}_{}\,,\,\bar d_m^{}\,]~=~\frac{i}{2}\,\bar d^{}_m\ .
\ee
The dynamical Lorentz generators are the boosts, given by 

\bea
j^-_{}&=&i\,x\,\frac{\partial\bar\partial}{\partial^+_{}} ~-~i\,x^-_{}\,\partial~+~i\,\Big( \theta^p_{}\bar\partial^{}_p\,-\lambda-1\Big)\frac{\partial}{\partial^+_{}}\,\ ,\cr 
\bar j^-_{}&=&i\,\bar x\,\frac{\partial\bar\partial}{\partial^+_{}}~ -~i\,x^-_{}\,\bar\partial~+~ i\,\Big(\bar\theta_p^{}\partial_{}^p+\lambda-1\,\Big)\frac{\bar\partial}{\partial^+_{}}\,\ .
\eea
They also preserve chirality because 

\be
[\,j^{-}_{}\,,\,d^m_{}\,]~=~\frac{i}{2}\,d^m_{}\,\frac{\partial}{\partial^+_{}}\ ,\qquad [\,j^{-}_{}\,,\,\bar d_m^{}\,]~=~\frac{i}{2}\,\bar d_m^{}\,\frac{\partial}{\partial^+_{}}\ ,
\ee
and satisfy the Poincar\'e algebra. In particular

\be
[\,j_{}^-\,,\,\bar j^+_{}\,]~=~-i\,j^{+-}_{}-j\ ,\qquad [\,j^-_{}\,,\,j^{+-}_{}\,]~=~i\,j^{-}_{}\ .
\ee
Half of the  supersymmetry generators 

\be
q^{\,m}_{\,+}=-{\partial^m}\,+\,\frac{i}{\sqrt 2}\,{\theta^m}\,{\partial^+}\ ;\qquad{{\bar q}_{\,+\,n}}=\;\;\;{{\bar \partial}_n}\,-\,\frac{i}{\sqrt 2}\,{{\bar \theta}_n}\,{\partial^+}\ ,
\ee
are kinematical. They satisfy

\be
\{\,q^{\,m}_{\,+}\,,\,{{\bar q}_{\,+\,n}}\,\}\,=\,i\,{\sqrt 2}\,{{\delta^m}_n}\,{\parp}\ ,
\ee
and anticommute with the chiral derivatives

\bea
\{\,q^{\,m}_{\,+}\,,\,{{\bar d}_n}\,\}\,=\,\{\,{d^m}\,,\,{{\bar q}_{\,+\,n}}\,\}\,=\,0\ .
\eea
The other half are dynamical, obtained by boosting the kinematical supersymmetries

\be
{q}^m_{\,-}~\equiv~i\,[\,\bar j^-_{}\,,\,q^{\,m}_{\,+}\,]~=~\frac{\bar\partial}{\partial^+_{}}\, q^{\,m}_{\,+}\ ,\qquad 
{\bar{q}}_{\,-\,m}^{}~\equiv~i\,[\, j^-_{}\,,\,\bar q_{\,+\,m}^{}\,]~=~\frac{\partial}{\partial^+_{}}\, \bar q_{\,+\,m}^{}\ .
\ee
These are the ``square-roots" of the light-cone Hamiltonian, in the sense that

\be
\{\, {q}^m_{\,-}\,,\,{\bar{q}}_{\,-\,n}^{}\,\}~=~i\,\sqrt{2}\,\delta^{\,m}_{~~n}\,\frac{\partial\bar\partial}{\partial^+_{}}\ .
\ee
All  operators  have been constructed so as to be Hermitian with respect to the quadratic form

\be
(\,\phi\,,\,\xi\,)~\equiv~2i\int d^4\!x\, d^4\!\theta \, d^4\,{\bar\theta}\,\,{\bar\phi}\,\frac{1}{\parp}\xi\ ,\label{inner}
\ee
where $\phi$ and $\xi$ are chiral superfields. In the fully interacting theory, the kinematical superPoincar\'e generators do not change, still acting linearly on the superfields, but the dynamical generators  act non-linearly, as we shall discuss in section $4$.


\subsection{Superconformal Algebra}
 In this section, we complete the construction of the $PSU(2,2\vert\, 4)$ generators.  Our procedure is to start from the simplest  kinematical generator in the conformal algebra, and generate the rest by commutation. 

The easiest starting point is  the  ``plus" component of the conformal transformations,

\bea
K^+\,=\,2i\,x\,{\bar x}\,\parp\ .
\eea
For ease of algebra, we continue to work at $x^+\,=\,0$, where $K^+$ is kinematical. Since we already know $j^{+-}$, the commutation relation

\bea
[\,K^{+},\,{p}^{-}\,]=-2i\,D+2i\,j^{+-}\ ,
\eea
yields  the dilatation generator,

\bea
D\,=\,i\,\left(\,x^-\parp-\,x{\bar \partial}\,-\,{\bar x}\partial-\,\frac{1}{2}\theta\frac{\partial}{\partial \theta}\,-\,\frac{1}{2}{\bar \theta}\frac{\partial}{\partial\bar{\theta}}\,\right)\ ,
\eea
which satisfies

\bea
[x\,,\,D]\,=\,i\,x\ ;\qquad [d^m\,,\,D]\,=\,-i\frac{1}{2}d^m\ . 
\eea
By boosting $K^+$, we obtain the kinematical generators $K$ and $\bar{K}$,

\bea
K\!\!\!&=&\!\!\!i \,[\, j^{-}, K^+\,]=2ix\left(x^-\parp-x{\bar \partial}-\theta\frac{\partial}{\partial\,\theta}+\lambda
\right)\!,\\
\bar{K}\!\!\!&=&\!\!\!i\,[\, \bar{j}^{-}, K^+\,]=2i{\bar x}\left(x^-\parp-{\bar x}\partial-{\bar \theta}\frac{\partial}{\partial {\bar \theta}}-\lambda
\right)\ ,
\eea
where $\lambda$ is the helicity counter of the previous section. These  generators do not change the chirality of the superfields on which they act, since

\bea
[\,K,\,d^m\,]~=~0\ ;\qquad[\,{\bar K},\, d^m\,]\, = \, 2i\,\bar{x}\,d^m\ .
\eea
The supersymmetry generators are now augmented by new ``conformal supersymmetries", easily obtained from their normal counterparts by

\bea
[\,K^+\,,\,q^m_-\,]\,=\,-\,\sqrt 2\,(\,i\,\sqrt 2\,{\bar x}\,{q^m_+}\,)\,=\,-\,\sqrt 2\,s^m_+\ .
\eea
These, together with their complex conjugates,

\bea
[\,K^+\,,\,{\bar q}_{-\,n}\,]\,=\,\sqrt 2\,(\,-\,i\,\sqrt 2\,x\,{\bar q}_{+\,n}\,)\,=\,\sqrt 2\,{\bar s}_{+\,n}\ ,
\eea
are kinematical. We note that $K^+$ and these supercharges should contain a dynamical term, but it is  multiplied by the parameter $x^+$, which we have set to zero without loss of generality.  

In a similar fashion, the dynamical conformal  supersymmetries are obtained by boosting  

\bea
\begin{split}
s^m_-\,=&\,i\,[\,j^-\,,\,s^m_+\,]\,\\
=&\,i\,\sqrt 2\,\left(\,x^-\,\parp\,-\,x\,\bar \partial\,-\,\theta\,\frac{\partial}{\partial\,\theta}\,
+\lambda+1
\right)\,\frac{1}{\parp}\,q^m_+\ ,\\
{\bar s}_{-\,n}\,=&\,i\,[\,{\bar j}^-\,,\,{\bar s}_{+\,n}\,]\,\\
=&\,-\,i\,\sqrt 2\,
\left(\,x^-\,\parp\,-\,\bar x\,\partial\,
-\,{\bar \theta}\,\frac{\partial}{\partial\,{\bar \theta}}\,
-\lambda+1
\right)\,\frac{1}{\parp}\,{\bar q}_{+\,n}\ .
\end{split}
\eea

\noindent Like the  supersymmetry transformations, the conformal supersymmetry generators $s$ act as square roots of the  conformal translations. Using these expressions, we  verify closure of the algebra by checking that the anticommutators

\bea
\{s^m_+,~\bar{s}_+{}_n\}= \sqrt{2}\delta^m_n K^+\ ; \quad 
\{s^m_+,~\bar{s}_-{}_n\}= \sqrt{2}\delta^m_n \bar{K}\ ;  \quad
\{s^m_-,~\bar{s}_+{}_n\}= \sqrt{2}\delta^m_n {K}\ .
\eea		
yield  exactly the same expressions as previously obtained for $K^+$, $K$ and $\bar K$. The superconformal algebra is completed by calculating the  dynamical conformal generator $K^{-}$  from 

\be
K^-\,=\,i\,[\,{\bar j}^-\,,\,K\,]\ ,
\ee
with the result

\bea
\begin{split}
K^-\,
=&\,2\,i\,\left(\,x^-\,\parp\,-\,{\bar x}\,\partial\,-\,{\bar \theta}
\frac{\partial}{\partial\,{\bar \theta}}\,
-\lambda +1
\right)\,\\
&\times\left(\,x^-\,\parp\,-\,x\,\bar \partial\,-\,\theta \frac{\partial}{\partial\,\theta}\,
+\lambda +1
\right)\,\frac{1}{\parp}\ .
\end{split}
\eea
The consistency of this expression is  checked  by calculating the  anticommutator,

\bea
\sqrt 2\,\delta^m_n\,K^-\,=\,\{\,s^m_-\,,\,{\bar s}_{-\,n}\,\}\ .
\eea
This  completes the construction of the superconformal algebra at the free level. We note for completeness a host of commutation relations  in light-cone form

\bea
\begin{split}
\{\,q^m_+\,,\,{\bar q}_{+n}\,\}~=~-\,{\sqrt 2}\,\delta^m_n\,p^+ \ ;\qquad \{q^m_-\,,\,{\bar q}_{-n}\,\}\,=\,-\,{\sqrt 2}\,\delta^m_n\,p^- \\
\,[\,s_{+}^{m}\,,\,p\,]~=~-\sqrt{2}{q}_{+}^{m} \ ;~~~~~~~~~~~~~\qquad [\,s_{-}^{m}\,,\,{\bar p}\,]~=~\sqrt{2}q_-^{m} \\
\{s^m_-\,,\,\bar{s}_-{}_n\}~=~\sqrt{2}\delta^m_n K^- \ ;\qquad \{s^m_-\,,\,\bar{s}_+{}_n\}~=~\sqrt{2}\delta^m_n\,K \\
\,[q_{-}^{m}\,,\,K]~=~\sqrt{2}\,s_{-}^{m} \ ;\qquad [\,q_{+}^{m}\,,\,K^{-}~]=~-\sqrt{2}\,s_- ^{m}\ .
\end{split}\eea
Finally, the anticommutators of the conformal and normal  supersymmetries  yield  new  operators, $J^m_{\,\,\,n}$, 

\be
\{q_{+}^{m}\,,\,\bar{s}_{-}{}_{n}\}=-i\delta^m_n(D+j^{+-}+ij)+2 J^{m}{}_{n}\ ,\ee
which generate  the $SU(4)$ Lie algebra

\be
[\,J^{m}{}_{n}\,,\,J^{p}{}_{q}\,]~=~\,\delta^{m}{}_{q}\,J^{p}{}_{n}-\,\delta^{p}{}_{n}\,J^{m}{}_{q}\ ,
\ee
and commute with the $SO(4,2)$ generators. The bosonic generators we have constructed generate 
 \bea
SO(4,2)\,\times\,SU(4)\;\sim\;SO(4,2)\,\times\,SO(6)\ ;
\eea
together with the fermionic generators, they form  the entire $PSU(2,2\,|\,4)$ algebra.

\renewcommand{\theequation}{4.\arabic{equation}}
\setcounter{equation}{0}
\section{Non-linear Realizations}
The superPoincar\'e algebra contains  three types of {\em dynamical} transformations, those generated by the light-cone Hamiltonian $p^-$, by the boosts $j^-$ and $\bar j^-$, and by the supersymmetries $q_-$ and $\bar q_-$.  

In the interacting theory, these transformations  act non-linearly on the superfields, while preserving the commutation relations of the Super-Poincar\'e algebra. The  form of these transformations  determines the fully interacting super-Poincar\'e invariant action. This decomposition is also true for the  conformal transformations contained in the full $PSU(2,2\,\vert\,4)$.

\subsection{Old Results}
Bengtsson {\em et al}~\cite{BBB} devised in $1982$ a systematic procedure for finding these non-linear transformations order by order in  the coupling constant $g$.  They expanded  the dynamical transformations of the fields as a power series in $g$, 

\bea
\begin{split}
\delta^{}_{p^-} \phi&= {\delta^0_{p^-}} \phi + {\delta^{g}_{p^-}} \phi  + {\delta^{g^2}_{p^-}} \phi+ \cdots\ ,\\
\delta^{}_{q_-} \phi&= {\delta^{0}_{q_-}} \phi + {\delta^{g}_{q_-}} + {\delta^{g^2}_{q_-}}\phi +\cdots\ , \\
\delta^{}_{j^-} \phi&= {\delta^{0}_{j^-}} \phi + {\delta^{g}_{j^-}} \phi +{\delta^{g^2}_{j^-}} \phi+\cdots\ ,
\end{split}
\eea
where the superscript denotes the order of the variation. Since the kinematical transformations remain unaltered  with no order $g$ corrections, much information is gained from the commutation relations  

\be
[\,{\delta^{}_j}\,,\,{\delta^{}_{p^-}}\,]\,\phi ~=~ 0\ ,\qquad [\,{\delta^{}_{j^{+-}}}\,,\,{\delta^{}_{p^-}} \,]\,\phi~=~\,i\,{\delta^{}_{p^-}}\phi\ ,
\label{CR1}\ee
(which determine the helicity and the number of $\partial^+$ in the variations)  as well as from requiring that 

\be
[\,\delta^{}_{\bar j^-}\,,\,\delta^{}_{p^-}\,]\,\phi ~=~ 0\ ,\label{CR2}
\ee
holds order by order in $g$. This allowed Bengtsson {\em et al} to determine these non-linear transformations to first order in $g$.  They also determined the non-linear dynamical supersymmetry transformations by boosting

\be
\delta^{}_{\bar q_{-}}\,\phi~=~i\,[\,\delta^{}_{\bar q_+},\,\delta^{}_{j^-}]\,\phi\ ,\qquad 
\delta^{}_{q_{-}}\,\phi~=~ i\,[\,\delta^{}_{q_+},\,\delta^{}_{\bar j^-}]\,\phi\ .
\ee
Due to algebraic complications, the authors did not proceed beyond the first order in coupling. 

In detail, their method relied on formulating  ans\"atze  for the order $g$ Hamiltonian variation,

\be
{\delta^g_{p^-}}\,{\phi}~=~-\,i\,g\,{\parp}^\mu\,[\,{{\bar \partial}^a}\,{\parp}^\rho\,{\phi}\,{{\bar
\partial}^b}\,{\parp}^\sigma\,{\phi}\,]\ ,
\ee
as well as for the boosts

\be
{\delta^g_{j^-}}\,{\phi}~=~-\,x\,\delta^g_{p^-}\,\phi + \delta^{g}_{spin}\,\phi\ ,
\ee
where the latter is the ``spin" change. 

Requiring closure of the commutators (\ref{CR1}) to order $g$ yields, keeping in mind that the variations act  {\em only} on the superfields, 

\be
a\,+\,b\,=\,1\ ,\qquad \mu\,+\,\rho\,+\,\sigma\,=\,0\ .
\ee

A simple calculation showed that the order $g$ commutation relation (\ref{CR2})
could not be satisfied unless there were several fields that entered the variation antisymmetrically. This was the first indication of the gauge structure function $f^{abc}$, which implied  several fields, labeled  with extra indices $a,b,c$. They assumed that it was the completely antisymmetric 3-form. Once these assumptions are made, the vanishing of the commutator can be achieved for $\mu=-1$. 

To summarize, the variation that satisfies Poincar\'e invariance to order $g$ is then  

\be
{\delta^g_{p^-}}\,{\phi^a}~=~-\,i\,g\,f^{abc}\,\frac{1}{\parp}\,(\,{\bar \partial}\,{\phi^b}\,{\parp}\,{\phi^c})\ .
\ee
In a similar fashion, they obtain the non-linear contribution to the boosts  by requiring

\be
[\,{\delta^{}_{j^-}}\,,\,\delta_{p^-}^{}\,]\,{}^g\,\phi~=~0\ ,
\ee
which implies, after a lengthy calculation, that 

\bea
\delta^g_{j^-}\,\phi^a~=~-\,x\,\delta^{g}_{p^-}\,\phi^a\,+i\,g\,f^{abc}\,\frac{1}{\parp}\,\left\{ \,(\,{\theta}\,
\frac{\partial}{\partial\,\theta}\,-\,1\,)\,{\phi^b}\,{\parp}\,{\phi^c}\,)\,\right\}\ .
\eea
Finally, by boosting the kinematical supersymmetries, they obtained  the dynamical ones 

\bea
\delta^g_{\bar q_{-}}\,\phi^a \!\!\!&=&
-\,g\,{f^{abc}}\,{\frac {1}{\partial^+}}\,\left(\,{\frac {\partial}{\partial\,\theta}}\,{\phi^b}\,
{\partial^+}\,{\phi^c}\,\right)\ ,\\
\delta^g_{q_{-}}\,\phi^a\!\!\!&=& g\,{f^{abc}}\,{\frac {(d)^4}{48\,\partial^+{}^3}}\,\left(\,
{\frac {\partial}{\partial\,\bar\theta}}\,{\bar\phi^b}\,
{\partial^+}\,{\bar\phi^c}\,\right),
\eea
where $(d)^4\equiv\epsilon_{mnpq}\,d^md^nd^pd^q$. Note that these transformations  do not contain transverse derivatives. 
The authors used the variations to generate the cubic interaction vertex from the kinetic two-point function, but  did not extend their method to higher order in $g$. As a result, their procedure fell short of showing that $f^{abc}$ is a three-form which satisfies the Jacobi identity, and  of deriving the four-point function using algebraic means. In the next section, we complete their program. 


\subsection{New Results: Symmetries}
The method described in the previous section is very generic, and does not make use of supersymmetry. Yet in supersymmetric theories, the Hamiltonian is a derived concept (as if the square-root of time were taken). Furthermore, Bengtsson {\em et al}  had to make inspired guesses for {\em both} the Hamiltonian and boosts separately. 

In this paper, we restrict our attention to the ${\cal N}=4$ theory which has a much larger invariance group, namely $PSU(2,2\,\vert\,4)$. Unlike the superPoincar\'e symmetry, it is a {\it {simple}} Lie superalgebra. This means that  it suffices to know one bosonic  kinematical conformal transformation and the form of the non-linearly realized supersymmetry to reconstruct the whole algebra for the interacting case, and therefore the fully interacting classical action!

For that reason, we only have to determine the dynamical supersymmetry to order $g$ and higher. The construction proceeds in several steps. First we  show that considerations of chirality, dimensional analysis, proper helicity and simple commutators restrict the first order dynamical supersymmetry to be of the form

\be
\delta^g_{\bar q_-}\,\phi^a~=~-g\,f^{abc}_{}\,\frac{1}{\partial^{+\,(2\nu+1)}}\left\{\,{\bar d}\,\partial_{}^{+\,\nu}\,\phi^b_{}\,\partial^{+\,(\nu+1)}_{}\,\phi^c\right\}\ .
\ee
Here 

\be
f^{abc}_{}~=~-f^{acb}_{}\ .
\ee
 $\bar d$  and $\bar q_+$ are interchangeable because of the antisymmetry of the structure function, and $\nu$ is a free parameter, to be fixed  by the algebra.  So as not to interrupt the flow of our arguments, the details of these calculations are relegated to  Appendix A.

We take the conjugate of this expression, and use the ``inside-out" relation to find

\be
\delta^g_{ q_-}\,\phi^a~=~-g\,f^{abc}_{}\,\frac{(d)^4_{}}{48\,\partial^{+\,(2\nu+3)}}\left\{\,{ d}\,\partial_{}^{+\,\nu}\,\bar\phi^b_{}\,\partial^{+\,(\nu+1)}_{}\,\bar\phi^c\right\}\ .
\ee
The next step is to evaluate  the anticommutator

\be
\{\,\delta^{}_{ q_-^m}\,,\,\delta^{}_{\bar q_{-\,n}}\,\}^g_{}\,\phi^a~=~-\sqrt{2}\,\delta^m_n\,\delta^g_{p^-}\,\phi^a\ ,
\ee
to first order in $g$. Use of chirality and the relation

\be
\bar q_+~=~\bar d-i\sqrt{2}\,\bar\theta\,\partial^+_{}\ ,
\ee
lead to 

\bea
\delta^{}_{p^-}\,\phi^a_{}~=~ -i\,\frac{\partial\bar\partial}{\partial^+}\,\phi^a_{} -\,i\,g\,f^{abc}\,\biggl\{\,\frac {1}{\partial^{+\,(2\nu+1)}}\,(\,{\bar
\partial}\,\partial^{+\,(\nu)}\phi^b\,\partial^{+\,(\nu+1)}\phi^c\,)\nonumber&& \\+\,\frac{(d)^4}{48\,{\parp}^{(2\nu+3)}}\,(\,{\partial}\,\partial^{+\,(\nu)}\,{\bar\phi}^b\,\partial^{+\,(\nu+1)}\,{\bar \phi}^c\,)\,\biggr \}+{\cal O}(g^2)\ .
\eea
This is of course the light-cone Hamiltonian. In Appendix A, we show that the dynamical supersymmetry variation does not extend beyond order $g$. Thus the classical Hamiltonian extends only up to order $g^2$. It is too cumbersome here to derive it  from the anticommutator, and  we will obtain it in a much simpler way from the action.

Because $\bar K$ is kinematical, void of order $g$ corrections, we can now {\em derive} the form of the non-linear boosts using the conformal group commutator

\be
[\,\delta^{}_{\bar K}\,,\,\delta^{}_{p^-}\,]\,\phi^a~=~-2\,i\,\delta^{}_{\bar j^-}\,\phi^a\ .
\ee
A straightforward computation yields 

\be
\delta^{g}_{\bar j^-}\,\phi^a~=~-\,\bar x\,\delta^g_{p^-}\,\phi^a\,-\,\frac{i\,g\,(d)^4\,f^{abc}}{48\,\partial^{+\,(2\nu+3)}}\left\{\,
(\bar\theta\frac{\partial}{\partial\bar\theta}+\nu-1)\partial^{+\,\nu}_{}\bar\phi^b\,\partial^{+\,(\nu+1)}_{}\bar\phi^c\right\}\ ,
\ee
neglecting the order $g^2$ contributions. We are now in a position to verify that

\be
[\,\delta^{}_{\bar j^-}\,,\,\delta^{}_{p^-}\,]\,\phi^a~=~0\ .
\ee
Evaluating this commutator to order $g$ yields terms proportional to $\nu$, such as 

\bea
[\,\delta^{}_{\bar j^-}\,,\,\delta^{}_{p^-}\,]^g\,\phi^a&=&
\frac{\nu\,g\,f^{abc}_{}}{\partial^{+(2\nu+2)}_{}}\,
\left\{ \partial^+_{}(\partial^{+(\nu-1)}_{}\bar\partial_{}^2\,\phi^b_{}\partial^{+(\nu+1)}\phi^c_{})~-\right.\cr
&&~~~~~~~~\left.-~2\,\bar\partial(\bar\partial\partial^{+\,\nu}\phi^b_{}\partial^{+(\nu+1)}_{}\phi^c_{}\,)\right\}+\cdots\ ,
\eea
which require $\nu=0$. Hence by using the superconformal algebra and chirality we have arrived at the unique realization of the $PSU(2,2\,\vert\,4)$ algebra. 

The procedure to obtain the remaining transformations is straightforward. In particular, the superconformal transformations are obtained through the commutator,

\bea
\delta^g_{\bar{s}_{-}}\phi^a\, &=&\,
\frac{1}{\sqrt{2}}\left[\delta^g_{\bar{q}_{-}},\,\delta_{\bar K} \right]\phi^a
={\sqrt{2}} i \bar{x}\delta^g_{\bar{q}_-}{\phi}^a \nonumber\\
&= &  {\sqrt{2}} i\,{\bar x}\,g\,{f^{abc}}\,\frac {1}{\parp}\left(\,
\frac {\partial}{\partial\,\theta}\,\phi^b\,{\parp}\,{\phi^c}\right )\ .
\eea
The total antisymmetry of the $f^{abc}$ and the Jacobi identities are obtained by requiring closure of the algebra. For instance, we calculate  the conformal generator $K^-$ in two independent ways,  from the commutator 

\be
\delta^{}_{K^-}~=~i\,[\,\delta^{}_{\bar j^-}\,,\,\delta^{}_{K}\,]\ ,\ee
or from the anticommutator

\be
\delta^{}_{K^-}~=~\frac{1}{4\sqrt{2}}\,\{\,\delta^{}_{s^m_-}\,,\,\delta^{}_{\bar s^{}_{-\,m}}\,\}\ .
\ee
Matching these two equations yields the Jacobi identity for the structure constants. In this gauge, space-time and internal symmetries are inextricably linked; conformal invariance requires the gauge symmetry of Yang-Mills theories.

Finally we note that the full dynamical supersymmetry operation can be written in the form

\be
\delta^{}_{\bar q_-}\,\phi^a_{}~=~\frac{1}{\partial^+}\,\left\{\,(\bar\partial\,\delta^{ab}_{}-gf^{abc}_{}\partial^+_{}\,\phi^c\,)\,\delta^{}_{\bar q_+}\,\phi^c\,\right\}\ ,\ee
suggesting  a covariant derivative structure

\be
{\cal D}^{ab}_{}~=~\bar\partial\,\delta^{ab}_{}-gf^{abc}_{}\partial^+_{}\,\phi^c\ ,\ee
although there are no known symmetries beyond the superconformal symmetry. 

\subsection{The Hamiltonian}
In this section, we show how to use these transformations to derive the fully interacting Hamiltonian starting from just its kinetic term. The action is of course well known, so that we will learn nothing new from this procedure except a methodology we expect to apply to other problems. 

It is easy to obtain the Hamiltonian from the light-cone action for ${\cal N}=4$ SuperYang-Mills, Eq.(\ref{ACTION})

\bea\label{hamil}
H\!\!\!&=& \!\!\!\int{d^4}x\,{d^4}\theta\,{d^4}{\bar \theta}\,\left\{
\bar\phi^a_{}\,\frac{2\partial\bar\partial}{\partial^{+2}}\,\phi^a_{}
~-\frac{4}{3}\,g\,f^{abc}_{}\,\Big(\frac{1}{\partial^+_{}}\,\bar\phi^a_{}\,\phi^b_{}\,\bar\partial\,\phi^c_{}+\frac{1}{\partial^+_{}}\,\phi^a_{}\,\bar\phi^b_{}\,\partial\,\bar\phi^c_{}\Big)
\right.\cr 
&&\left.+\,g^2f^{abc}_{}\,f^{ade}_{}\Big( \frac{1}{\partial^+}
(\phi^b\,\partial^+\phi^c)\frac{1}{\partial^+_{}}\,
(\bar\phi^d_{}\,\partial^+_{}\,\bar\phi^e)
+\frac{1}{2}\,\phi^b_{}\bar\phi^c\,\phi^d_{}\,\bar\phi^e\Big)\right\}\ ,\\
&\equiv&{H}_{}^0~+~{H}_{}^g~+~{ H}_{}^{g^2}\ .
\eea
We require that supersymmetry variations leave the Hamiltonian invariant,

\be
\delta^{}_{\bar q_-} H~=~0\ .
\ee
Expanding  in the coupling constant $g$ leads to three conditions

\bea
&&\delta^0_{\bar q_-}\,{{ H}^0}\;=\;0\ , \label{h0}\\
&&\delta^g_{\bar q_-}\,{{ H}^0}\,+\,{\delta^0_{\bar q_-}}\,{{ H}^g}\;=\;0\ , \label{h1}\\
&&{{\delta^g_{\bar q_-}}}\,{{ H}^g}\,+\,{{\delta^0_{\bar q_-}}}\,{{H}^{g^2}}\;=\;0\ ,\label{h2}
\eea
which offer a systematic procedure for finding $H^g$ and then $H^{g^2}$, starting from $\delta_{\bar q_-}$ and the free Hamiltonian $H^0$. 

The  lowest order condition trivially vanishes, but we can use the second to infer the form of $H^g$. We first evaluate 

\bea
\delta^g_{\bar q_-}\,{{ H}^0}~=~\delta_{q^{\,-}}^g\,{\biggl \{}\,\int\,\,{{\nbar \phi}^a}\,{\frac {2\,{\partial}{\bar \partial}}{{\parp}^2}}\,{\phi^a}\,{\biggr \}}\ .
\eea
A series of simple algebraic steps which use integration by parts of both $d$ and $\partial^+$, as well as chirality, lead to 
 
\bea
\delta_{q_{\,-}}^g\,{\biggl \{}\,\int\,{{\nbar \phi}^a}\,{\frac {2\,{\partial}{\bar \partial}}{{\parp}^2}}\,{\phi^a}\,{\biggr \}}\,=\,2\,g\,{f^{abc}}\,\int\,{{\nbar \phi}^b}\,d\,{{\nbar \phi}^c}\,{\frac {{\partial}\,{\bar \partial}}{{\parp}^2}}\,{\phi^a}\ .
\eea
Using Eq. (\ref{h1}), we obtain 

\be
{\delta^0_{\bar q_-}}\,{{ H}^g}~=~-2\,g\,\int\,{f^{abc}}\,{{\nbar \phi}^b}\,d\,{{\nbar \phi}^c}\,{\frac {{\partial}\,{\bar \partial}}{{\parp}^2}}\,{\phi^a}\ ,
\ee
which tells us the general structure of  $H^g$. Rather than directly rewriting this expression as the sum of three variations, one on each superfield, we consider  the variation

\bea
{{\delta^0_{q_{\,-}}}}\,{\biggl \{}\,\,g\,{f^{abc}}\,\int\,{\frac {1}{\parp}}\,{\phi^a}\,{{\nbar \phi}^b}\,{\partial}\,{{\nbar \phi}^c}\,{\biggr \}}\ ,
\eea
which yields three terms, (one of which is trivially zero),

\bea
g\,{f^{abc}}\,\int\,{\frac {1}{\parp}}\,{\phi^a}\,{\frac {\bar \partial}{\parp}}\,d\,{{\nbar \phi}^b}\,{\partial}\,{{\nbar \phi}^c}
+\,g\,{f^{abc}}\,\int\,{\frac {1}{\parp}}\,{\phi^a}\,{{\nbar \phi}^b}\,{\frac {{\partial}\,{\bar \partial}}{\parp}}\,d\,{{\nbar \phi}^c}\ .
\eea
and express  the first term in the above expression in two different ways. One is to act ${\frac {\parp}{\parp}}$ on the first $\phi^a$ and integrate by parts the $\parp$; the other is to use duality on  ${\nbar \phi}^c$, followed by partial integrations on both $\bar d$ and $\partial^+$.   Comparing the two resulting  expressions yields   

\bea
g\,{f^{abc}}\,\int\,{\frac {1}{\parp}}\,{\phi^a}\,{\frac {\bar \partial}{\parp}}\,d\,{{\nbar \phi}^b}\,{\partial}\,{{\nbar \phi}^c}\,
=\,{\frac {1}{2}}\,{\biggl \{}\,g\,{f^{abc}}\,\int\,{\frac {1}{\parp}}\,{\phi^a}\,{{\nbar \phi}^b}\,{\frac {{\partial}\,{\bar \partial}}{\parp}}\,d\,{{\nbar \phi}^c}\,{\biggr \}}\ .
\eea
Hence the variation

\bea
\delta^0_{q_{\,-}}\,{\biggl \{}\,\,g\,{f^{abc}}\,\int\,{\frac {1}{\parp}}\,{\phi^a}\,{{\nbar \phi}^b}\,{\partial}\,{{\nbar \phi}^c}\,{\biggr \}}\,
=\,\frac{3}{2}\,g\,{f^{abc}}\,\int\,{\frac {1}{\parp}}\,{\phi^a}\,{{\nbar \phi}^b}\,{\frac {{\partial}\,{\bar \partial}}{\parp}}\,d\,{{\nbar \phi}^c}\ ,
\eea
leading to the already known three-point function. One can also show that the variation of the complex conjugate  of the 3-point vertex vanishes

\bea
\delta^0_{q_{\,-}}\,{\biggl \{}\,\,g\,{f^{abc}}\,\int\,{\frac {1}{\parp}}\,{{\nbar \phi}^a}\,{\phi^b}\,{\bar \partial}\,{\phi^c}\,{\biggr \}}~=~0\ .
\eea

Having obtained the three-point function, we can now vary it to generate the four-point function, using Eq.(\ref{h2}). 
 The full three-point function has two parts, one that contains the transverse derivative $\partial$, and its complex conjugate  that contains $\bar\partial$.  However, we know that $H^{g^2}$ does not contain any transverse derivatives, since it is obtained from the supersymmetries at order $g$ which have  no transverse derivatives. Hence  consistency requires that

\bea\label{cons}
\delta^g_{q_-}\,H^g_\partial ~=~0\label{delpart}\ ,
\eea
since $\delta^0_{q_-}$ contains no $\partial$.  The proof proceeds in two steps. We first show that 

\be\label{del}
\delta^g_{q_-}\,H^g_\partial ~=~i\,\int\frac{1}{\partial^+}\phi^a\left.[\,\delta^g_{q_-},\,\delta^g_{p^-}\,]\,\bar\phi^a\,\,\right|_{\mbox{$\partial$-part}}\ ,
\ee
after many algebraic manipulations detailed in Appendix B. 

The algebraic requirement that the supersymmetries  commute with the Hamiltonian, implies to order $g^2$ that 

\be\label{algerq}
[\,\delta^0_{q^-}\,,\,\delta^{g^2}_{p^-}\,]\,+\,[\,\delta^g_{q^-}\,,\,\delta^g_{p^-}\,]~=~0\ .
\ee
The Hamiltonian variation contains both  $\partial$ and $\bar\partial$ parts, while $\delta^0_{q^-}$ involves only $\bar\partial$. Hence this equation breaks up into two separate equations. The first is 

\bea
\left.[\,\delta^g_{q^-}\,,\,\delta^g_{p^-}\,]\,\right|_{\mbox{$\partial$}}~=~0\ ,\eea
which as we show in the same Appendix, is satisfied as long as  the structure functions are totally antisymmetric and obey the Jacobi identity. This satisfies the consistency requirement (\ref{cons}). The second relation 

\be
\left.[\,\delta^0_{q^-}\,,\,\delta^{g^2}_{p^-}\,]\,+\,[\,\delta^g_{q^-}\,,\,\delta^g_{p^-}\,]\right|_{\mbox{$\bar\partial$}}~=~0\ ,
\ee
can now be used to extract the form of $H^{g^2}$ from  Eq.(\ref{h2}). The three-point variation $\delta^g_{q^-} H^g_{\bar\partial}$ can be expressed as  

\bea
\delta^g_{q_-} H^g_{\bar\partial} &=& i \int \frac{1}{\partial^+}\bar\phi^a \,[\,\delta^g_{q_-}\,,\,\delta^g_{p^-}\,]\, \phi^a \Bigl|_{\mbox{$\bar\partial$-part}} \nonumber\\
&=&-i \int \frac{1}{\partial^+}\bar\phi^a \,[\,\delta^0_{q_-}\,,\,\delta^{g^2}_{p^-}\,]\, \phi^a\nonumber\\
&=&-i \int\left\{ \frac{1}{\partial^+}\bar\phi^a \delta^0_{q_-}\delta^{g^2}_{p^-}\phi^a - \frac{1}{\partial^+}\bar\phi^a\delta^{g^2}_{p^-}\frac{\bar\partial}{\partial^+}{q_+}\phi^a\right\} \nonumber\\
&=& -i \int\left\{ \frac{1}{\partial^+}\bar\phi^a \delta^0_{q_-}\delta^{g^2}_{p^-}\phi^a +
\frac{1}{\partial^+}\delta^0_{q_-}\bar\phi^a\delta^{g^2}_{p^-}\phi^a \right\}\nonumber\\
&=&-\delta^0_{q^-}\left\{\, i\int\frac{1}{\partial^+} \bar\phi^a\delta^{g^2}_{p^-}\phi^a\, \right\}\ ,
\eea
where we have used Eq.(\ref{algerq}). Hence the four-point function,

\bea
H_{}^{g^2}
~=~i\int\frac{1}{\partial^+} \bar\phi^a\,\delta^{g^2}_{p^-}\phi^a~=~-\frac{i}{4\sqrt{2}} \int \frac{1}{\partial^+} \bar\phi^a \,\{ \,\delta^g_{q^-}\,,\,\delta^g_{\bar q^-}\,\}\,\phi^a\ .
\eea
This completes the construction of the classical Hamiltonian. 

\renewcommand{\theequation}{5.\arabic{equation}}
\setcounter{equation}{0}
\section{Hamiltonian as a Quadratic Form}
Our algebraic formulation of the ${\cal N}=4$ SuperYang-Mills theory enables us to write its Hamiltonian in a particularly suggestive form. We note that  the free Hamiltonian

\bea
H^0_{}~=~\int{d^4}x\,{d^4}\theta\,{d^4}{\bar \theta}\,
\bar\phi^a_{}\,\frac{2\,\partial\bar\partial}{\partial^{+2}}\,\phi^a_{}\ ,
\eea
can be rewritten as a quadratic form

\bea
H^0_{}~=~\frac{1}{2\,\sqrt{2}}\,(\,{\mathcal W}_0\,,\,{\mathcal W}_0\,)\ ,
\eea
using the inner product notation of Eq.(\ref{inner}),  where

\bea 
{\mathcal W}^a_{0}~=~ \frac{\partial}{\partial^+} {\bar q}^{}_{+}\phi_{}^a\ ,
\eea
is a fermionic superfield, the {\em free} dynamical supersymmetry variation of the superfield ($SU(4)$  spinor indices are summed over).  The proof is straightforward, and requires  integration by parts and the use of of the inside-out property of the superfields. 

In order to generalize this simple formula to the fully interacting Hamiltonian, we note that the three-point function can be expressed in the suggestive form

\bea
\frac{4}{3}\,g\,f^{abc}\int\frac{1}{\partial^+}\phi^a{\bar \phi}^b\partial{\bar \phi}^c=
\frac{i}{\sqrt2}\,g\,f^{abc} \int \frac{\partial}{\partial^+{}^2}\phi^a {\bar q}_+{}_m\frac{1}{\partial^+} (d^m {\bar \phi}^b \partial^+{\bar \phi}^c)\ .\label{identity}
\eea
The detailed proof of this identity is left to Appendix B. Similarly, the two terms that describe the the four-point interaction can be rewritten as

\bea
\begin{split}
&\int g^2f^{abc}_{}\,f^{ade}_{}\Big\{\,\frac{1}{\partial^+_{}}(\phi^b\,\partial^+\phi^c)\frac{1}{\partial^+_{}}\,(\bar
\phi^d_{}\,\partial^+_{}\,\bar\phi^e)+\frac{1}{2}\,\phi^b_{}\bar\phi^c\,\phi^d_{}\,\bar\phi^e\Big\} \\
&~~~=\int
\frac{i}{\sqrt{2}} g^2f^{abc}f^{ade}\frac{1}{\partial^+}({\bar d}_m\phi^b\partial^+\phi^c)\frac{1}{\partial^+{}^2}
(d^m{\bar \phi}^d\partial^+{\bar \phi}^e)\ ,
\end{split}
\eea
also proved in the same Appendix. 

With their help, it is easy to see that the fully interacting Hamiltonian~(\ref{hamil}) can be expressed as a quadratic form

\bea
H&=&\!\!\frac{1}{2\sqrt{2}}\, \left(\,{\mathcal W}^a_{}\,,\,{\mathcal W}^a_{}
\,\right)\ ,\label{square}
\eea
where now

\bea
{\mathcal W}^a_{}~=~ \frac{\partial}{\partial^+} {\bar q}_{+}\phi^a
-gf^{abc}\frac{1}{\partial^+} ({\bar d}\phi^b\partial^+\phi^c)\ ,
\eea
is the complete (classical) dynamical supersymmetry variation. The power of supersymmetry allows for this simple rewriting of the fully interacting Hamiltonian.  


\section{Conclusions}
We have used purely algebraic techniques to reconstruct the ${\cal N}=4$ Yang-Mills theory on the light-cone. Knowledge of the  dynamical supersymmetry transformations, which  are of first order in the coupling,  suffice to fix the full symmetry of the theory, and to write the Hamiltonian as a quadratic form. The simple answers we found in this exercise suggest several lines of inquiry which we are presently pursuing. 

We conjecture that, in the quantum theory,  the structure of the Hamiltonian remains a quadratic form, except that ${\cal W}^a$ picks up quantum corrections of order $\hbar$ and higher. 

As the full quantum action is generated by the Dirac-Feynman path integral, this simple quadratic form suggests that we seek a change of variables from the superfields $\phi^a$ to the fermionic superfields ${\cal W}^a$.  In addition, we see that field configurations for which ${\cal W}^a=0$ have vanishing energy,   and their study should prove interesting.

Finally, we plan to apply the same techniques to ${\cal N}=8$ supergravity, and generate its classical action through the supersymmetry transformations. We intend to use these techniques to derive the hitherto unknown  four- and higher-point functions in terms of chiral superfields.

\section{Acknowledgements}
We wish to thank Steve Pinsky for suggesting writing the cubic vertex as a cross-term. The research of three of the authors (S. A, S. K. and P. R.) was supported in part by the US Department of Energy under grant DE-FG02-97ER41029. One of us (S.A.) also  acknowledges support from 
 a McLaughlin Fellowship from the University of Florida.

\newpage

\renewcommand{\theequation}{A-\arabic{equation}}
\setcounter{equation}{0}  
\section*{APPENDIX A : The Supersymmetry Variation}  

\vskip 1.5cm
\noindent In this Appendix, we present two results. First, we construct the supersymmetry variation to order $g$ based on various checks. Secondly, we show that it is not possible to build a variation at order $g^2$ with the requisite properties, thus proving that the supersymmetry variations in this theory do not extend beyond order $g$.

Before attempting to establish the form of the dynamical supersymmetry variation, we list the algebraic constraints on its structure. We start with the commutator involving the kinematical ${\bar j}^+$ and ${\bar q}_{+n}$,

\be
[\,\delta_{{\bar j}^+}\,,\,\delta_{{\bar q}_{-n}}\,]\,=\,-\,i\,\delta_{{\bar q}_{+n}}\ .
\ee
Kinematical generators do not involve any $g$-dependent contributions, thus implying that

\be
\label{one}
[\,\delta_{{\bar j}^+}\,,\,\delta_{{\bar q}_{-n}}^{(g\,,\,g^2)}\,]\,=\,0\ .
\ee

We deduce that $\delta_{{\bar q}_{-n}}^{(g\,,\,g^2)}$ cannot have a $\partial$, since ${\bar j}^+$ contains an $\bar x$. We also note that,

\be
\label{two}
[\,\delta_{j^+}\,,\,\delta_{{\bar q}_{-n}}^{(g\,,\,g^2)}\,]\,=\,0\ ,
\ee
which (by the same argument) rules out the possibility of $\bar \partial$  in the variation. Hence there are no transverse space derivatives in $\delta_{{\bar q}_{-n}}^{(g\,,\,g^2)}$.  

Turning to dimensional analysis, and keeping track of factors of $\hbar$, we find that 

\be
\label{appfive}
{\delta_{{\bar q}_-}}\,\phi\,\propto\,g^A\,\hbar^B\,\phi^C
\ee
subject to the (mass-)dimensional constraint

\be
\label{apptwo}
A\,-2\,B\,-\,C\,=\,-\,1\ .
\ee
Since the left-hand side of Eq.(\ref{appfive}) carries a lower spinor index, the right-hand side necessarily contains one of the following quantities:

\bea
\begin{split}
&{\bar d}_m\ ;\qquad {\bar q}_m\ ;\qquad {(d)}^3_m\,=\,\epsilon_{mnpq}\,d^n\,d^p\,d^q\ ;\qquad {\{q\,{(d)}^2\}}_m\,=\,\epsilon_{mnpq}\,q^n\,d^p\,d^q\ \\
&{\{d\,{(q)}^2\}}_m\,=\,\epsilon_{mnpq}\,d^n\,q^p\,q^q\ ;\qquad {(q)}^3_m\,=\,\epsilon_{mnpq}\,q^n\,q^p\,q^q\ . 
\end{split}
\eea
We ignore these expressions and any factors of $\parp$ when using dimensional analysis since they all have a zero mass dimension which is the quantity we have tracked to obtain Eq. (\ref{apptwo}). For the classical theory, we set $B=0$ and see that the term proportional to $g$ involves two superfields while the term proportional to $g^2$ contains three.

We now set $\hbar\,=\,1$ and list the various (length-)dimensions and helicities of variables that occur in this theory:

\begin{center}
\begin{tabular}{||c c c||}
\hline\hline
Variable & Helicity ($h$) & Dimension ($D$) \\
\hline
$\phi$ & $+1$ & $0$  \\
$\bar \phi$ & $-1$ & $0$  \\
$x$ & $+1$ & $+1$  \\
$\bar x$ & $-1$ & $+1$ \\
$\partial$ & $+1$ & $-1$ \\
$\bar\partial$ & $-1$ & $-1$ \\
$d^m$ & $+\,1/2$ & $-\,1/2$  \\
${\bar d}_n$ & $-\,1/2$ & $-\,1/2$ \\
$q^m$ & $+\,1/2$ & $-\,1/2$  \\
${\bar q}_n$ & $-\,1/2$ & $-\,1/2$ \\
\hline\hline
\end{tabular}
\end{center}
\vskip .5cm

From the lowest order variation,

\bea
{{\delta^0_{\bar q_{-\,m}}}}\,\,\phi\,=\,\frac{\partial}{\parp}\,{\bar q}_{+\,m}\,\phi\ ,
\eea
we infer that the dynamical supersymmetry variations have a length dimension of $-\,\frac{1}{2}$ and a helicity of $\frac{3}{2}$. Another requirement is that these variations respect the chirality of the superfield they act on. Indeed the lowest order variation, trivially commutes with the chiral derivatives.

\subsection*{Variation at order $g$}
Having established general guidelines we now study possible candidates for the variation at order $g$. First consider an Ansatz with two chiral superfields (and a derivative)

\be
\label{appone}
\delta^{\,\,g}_{{\bar q}_{-\,m}}\,\phi\,\propto\,{\bar d}_m\,\phi\,\phi\ ,
\ee
and ask that it leave chirality invariant. In other words, we want

\be
\{\,\delta^{\,\,g}_{{\bar q}_{-\,m}}\,,\,d^n\,\}\,=\,0\ .
\ee
Since $\{{\bar d}_m\,,\,d^n\,\}\,=\,-\,i\,\sqrt 2\,\parp\,\delta_m^n$ we obtain

\be
\{\,\delta^{\,\,g}_{{\bar q}_{-\,m}}\,,\,d^n\,\}\,\phi\,=\,-\,i\,\sqrt 2\,g\,{\parp}\,\phi\,\phi\,\;\delta_m^n\ ,
\ee
which is non-zero. We are thus forced to introduce an antisymmetric function $f^{abc}$ into Eq. (\ref{appone}) and have the ${\bar d}_m$ act only on one $\phi$. We also introduce factors of $\parp$ into the relation to ensure that antisymmetry does not automatically render it zero,

\be
\delta^{\,\,g}_{{\bar q}_{-\,m}}\,{\phi^a}\,\propto\,f^{abc}\,\frac{1}{{\parp}^{(2\nu+1)}}\,{\biggl (}\,{\bar d}_m\,{\parp}^\nu\,\phi^b\,{\parp}^{(\nu+1)}\,\phi^c\,{\biggr )}\ ,
\ee
with $f^{abc}\,=\,-\,f^{acb}$. Note at this stage that our requirements are insufficient to prove antisymmetry between the $a$ \& $b$ indices or between $a$ \& $c$. This will be proven by algebraic means described in Appendix {\bf {B}}. The factor of $\frac{1}{{\parp}^{(2\nu+1)}}$ in the denominator balances the factors of $\parp$ introduced into the numerator. The anticommutator of this modified Ansatz with $d^n$ now reads,

\be
\{\,\delta^{\,\,g}_{{\bar q}_{-\,m}}\,,\,d^n\,\}\,\phi^a\,=\,-\,i\,\sqrt 2\,g\,f^{abc}\,{\parp}^{(\nu+1)}\,\phi^b\,{\parp}^{(\nu+1)}\phi^c\,\;\delta_m^n\,=\,0\ ,
\ee
by antisymmetry. This analysis thus leads us to the following form,

\be
\label{appfour}
\delta^{\,\,g}_{{\bar q}_{-\,m}}\,{\phi^a}\,\propto\,g\,f^{abc}\,\frac{1}{{\parp}^{(2\nu+1)}}\,{\biggl (}\,{\parp}^\nu{\bar d}_m\,\phi^b\,{\parp}^{(\nu+1)}\,\phi^c\,{\biggr )}\ . 
\ee
\vskip 0.3cm
Instead of beginning with Eq. (\ref{appone}), we could equally well have started with

\be
\delta^{\,\,g}_{{\bar q}_{-\,m}}\,\phi\,\propto\,{\bar q}_m\,\phi\,\phi\ ,
\ee
which is manifestly chiral,  but it ruins the anticommutator,

\be
\{\,\delta^{\,\,g}_{{\bar q}_{-\,m}}\,,\,\delta_{q_+^n}\,\}\,=\,0\ .
\ee
Restoring this anticommutator, requires the introduction of the antisymmetric
structure function and once again, leads to result (\ref{appfour}). Fixing the
value of $\nu$ (and the constant of proportionality) requires algebraic
computation and is presented in section {\bf 4}.

A second candidate for the variation at order $g$, is one proportional to $g\,\phi\,{\nbar \phi}$. This Ansatz has helicity $0$ and is manifestly non-chiral. It requires the introduction of three chiral derivatives to reach the target of $h\,=\,\frac{3}{2}$. However the inside out relation tells us that,

\be
{(d)}^3_m\,{\nbar \phi}\,\sim\,{\bar d}_m\,\phi\ ,
\ee
making this Ansatz proportional to the one in Eq. (\ref {appone}). 

Finally, we consider the totally antichiral Ansatz, proportional to $g\,{\nbar \phi}\,{\nbar \phi}$. Characterized by $h\,=\,-2$, it requires seven chiral derivatives to achieve a helicity of $\frac{3}{2}$. The inside-out relations again render this expression proportional to the very first Ansatz.

Thus at order $g$, the only viable Ansatz for the dynamical supersymmetry variation contains two chiral superfields and reads,

\be
\delta^{\,\,g}_{{\bar q}_{-\,m}}\,{\phi^a}~=~-\,g\,f^{abc}\,\frac{1}{{\parp}^{(2\nu+1)}}\,{\biggl (}\,{\parp}^\nu{\bar d}_m\,\phi^b\,{\parp}^{(\nu+1)}\,\phi^c\,{\biggr )}\ . 
\ee
\subsection*{Variation at order $g^2$}
The requirements on the supersymmetry variation at order $g^2$ are that it involve three superfields, have a length dimension of $-\,\frac{1}{2}$, a helicity of $\frac{3}{2}$ and contain no transverse derivatives. We offer a ``proof by exhaustion" that such an object does not exist.

We start by studying the case where the variation is proportional to three chiral superfields : ${g^2}\,\phi\,\phi\,\phi$. This expression has $D=0$ and $h\,=\,3\,$. The single lower spinor index is again introduced using,

\be
{\bar d}\ ;\qquad {\bar q}\ ;\qquad {(d)}^3\ ;\qquad {(d)}^2\,q\ ;\qquad d\,{(q)}^2\ ;\qquad {(q)}^3
\ee
These spinor expressions carry either helicities of $-\,\frac{1}{2}$ or $+\,\frac{3}{2}$ and when introduced into ${g^2}\,\phi\,\phi\,\phi$ result in a net helicity of $\frac{5}{2}$ or $\frac{9}{2}$. Achieving a net helicity of $\frac{3}{2}$ thus requires transverse derivatives (in this case $\bar \partial$). Since this would violate the commutator,

\bea
[\,j^+\,,\,{\bar q}_{-n}\,]\,=\,[\,x\,\parp\,,\,{\bar q}_{-n}\,]\,=\,0\ ,
\eea
the Ansatz is ruled out.

Other possible starting points involve mixtures of chiral and anti-chiral superfields. Consider a variation proportional to: ${g^2}\,\phi\,\phi\,{\nbar \phi}$ (this expression is manifestly non-chiral, something we will not worry about yet). The Ansatz has a helicity of $+1$ and $D=0$. Since we require a lower spinor index, we could introduce any of the following,

\be
\bar d\ ;\qquad {\bar q}\ ;\qquad {(d)}^3\ ;\qquad {q}\,{(d)}^2\ ;\qquad {(q)}^2\,d\ ;\qquad {(q)}^3\ .
\ee
The spinor expressions have helicities of either $-\,\frac{1}{2}$ (which must then be accompanied by a $\partial$) or $+\,\frac{3}{2}$ (which needs a $\bar \partial$). Either way, the Ansatz is not permitted. The other mixed Ansatz, proportional to ${g^2}\,\phi\,{\nbar \phi}\,{\nbar \phi}$ is ruled out by the same reasoning.

Finally, we consider the antichiral Ansatz: ${g^2}\,{\nbar \phi}\,{\nbar \phi}\,{\nbar \phi}$ with $h\,=\,-\,3\,$ and $D\,=\,0$. Helicity matching using just chiral derivatives turns the expression into one of the previous cases (since it would take nine chiral derivatives).

Thus any supersymmetry variation at order greater than $g$ necessarily contains one or more transverse space derivatives based on helicity and dimensional-requirements. The presence of a transverse derivative always ruins the commutator with either $j^+$ or ${\bar j}^+$ thus proving that supersymmetry variations in this theory, end at order $g$. The {\bf complete supersymmetry} variation is rather simple:

\bea
{\delta_{{\bar q}_-}}\,\phi^a\,=\,{\frac {1}{\parp}}\,{\biggl \{}\,(\,\partial\,\delta^{ab}\,-\,g\,f^{abc}\,{\parp}\,{\phi^c}\,)\,\delta^{}_{{\bar q}_+}\,{\phi^b}\,{\biggr \}}\ .
\eea
We note that it is always possible to alter a given Ansatz, using the index-less objects: $\epsilon_{mnpq}\,d^m\,d^n\,d^p\,d^q$ and $\epsilon^{mnpq}\,d_m\,d_n\,d_p\,d_q$. However, these simply act on the superfields, conjugating them (according to the inside-out relations) and thus reproduce one of the cases already covered. 

Our techniques are equally applicable to the larger $\mathcal N=8$ Supergravity theory in four dimensions. However, the dynamical supersymmetry variations in that theory pick up an infinite set of corrections (at every order in $\kappa$) making the theory itself more difficult to build. Order by order in $\kappa$, it is still easy to constrain the structure of the variation based on the various considerations discussed here. The supergravity variations do contain transverse derivatives and evade our arguments regarding Equations (\ref{one}) and (\ref{two}). This is due to the presence of two transverse derivatives in the action for that theory allowing structures wherein the contributions from the two derivatives cancel each other. In reference~\cite{ABR2} we discuss the possible form of the variations in the $\mathcal N=8$ theory at order $\kappa^2$.

\vskip 1.5cm


\setcounter{equation}{0}
\renewcommand{\theequation}{B-\arabic{equation}} 
\section*{ APPENDIX B : Mathematical Details}

\vskip 1.5cm
\subsection*{Explicit proof: $\delta^g_{q_-}(\mbox{3-pt function})\Bigl|_{\mbox{$\partial$-part}} 
\,=\,0$} 
The detailed proof for Eq.~(\ref{delpart}) proceeds in two steps. First we show that the supersymmetry variation on the $\partial$-dependent three-point function

\be
{ H}^g_{\;\partial}~\equiv~-\,\frac{4}{3}\,g\,f^{abc}\,\int\,\frac{1}{\parp}\phi^a\,{\bar \phi}^b\,\partial\,{\bar \phi}^c\ ,
\ee
can be expressed as

\be
\delta^g_{q_-}\,{ H}^g_{\;\partial}~=~2\,i\,\int\frac{1}{\partial^+}\phi^a\left.[\,\delta^g_{q_-},\,\delta^g_{p^-}\,]\,\bar\phi^a\,\,\right|_{\mbox{$\partial$-part}} \ .
\ee
we then show that

\be
\left. [\,\delta^g_{q_-},\,\delta^g_{p^-}\,]\,\bar\phi^a\,\,\right| 
_{\mbox{$\partial$-part}}~=~0\ ,
\ee
thus implying that  

\be
\delta^g_{q_-}\,{ H}^g_{\;\partial} \,=\,0\ .
\ee
The starting point is to rewrite the three-point function as the product of a chiral superfield and the order $g$ Hamiltonian variation,

\be
{ H}^g_{\;\partial}~=~
i\,\frac{4}{3}\int\frac{1}{\partial^+}\phi^a\delta^g_{p^-}\bar\phi^a\Bigl|_{\mbox{$\partial$}}\ .
\ee
The order $g$ supersymmetry variation on the three-point function is then 

\be
\delta^g_{q_-}\,{ H}^g_{\;\partial}~=~i\,\frac{4}{3}\int 
\left\{\frac{1}{\partial^+}\delta^g_{q_-}\phi^a\delta^g_{p^-}\bar\phi^a 
+\left.
\frac{1}{\partial^+}\phi^a\delta^g_{q_-}\delta^g_{p^-}\bar\phi^a\right\}\right|_{\mbox{$\partial$}} \ .
\ee
Work with the second term: expand $\delta^g_{p^-}\bar\phi$ explicitly and recombine terms, keeping $\delta^g_{q_-}\bar\phi$ explicit. This leads to

\bea
\begin{split}
&i\int\frac{1}{\partial^+}\phi^a\delta^g_{q_-}\delta^g_{p^-}\bar\phi^a\Bigl|_{\mbox{$\partial$-part}}\\
&=g\,f^{abc}\,\int\,\frac{1}{\partial^+{}^2}\phi^a\left[\,\partial\delta^g_{q_-}\bar\phi^b\partial^+\bar\phi^c\,+\,\partial\phi^b\partial^+\delta^g_{q_-}\bar\phi^c\, 
\right]\ ,\\
&=g\,f^{abc}\,\int\,\left\{\partial(\frac{1}{\partial^+}\phi^a\bar\phi^c)\delta^g_{q_-}\bar\phi^b\,+\,\partial^+(\frac{1}{\partial^+}\phi^a\partial\bar\phi^c)\delta^g_{q_-}\bar\phi^b\right\}\ ,\\
&=2\,g\,f^{abc}\,\int\,\frac{1}{\partial^+}\phi^a\partial\bar\phi^c\delta^g_{q_-}\bar\phi^b\ ,\\
&=2\,i\,\int \, \frac{1}{\partial^+} \delta^g_{q_-} \phi^a 
\delta^g_{p^-} \bar \phi^a \Bigl|_{\mbox{$\partial$-part}}\ ,
\end{split}
\eea
where we have made use of the inside-out relations in the last three steps. Thus, the variation of the three-point function can be 
written as

\be
\label{proof1}
\delta^g_{q_-}\,{ H}^g_{\;\partial}~=~2\,i\int 
\frac{1}{\partial^+}\phi^a\delta^g_{q_-}\delta^g_{p^-}\bar\phi^a\Bigl|_{\mbox{$\partial$-part}} \ .
\ee
The similarity of the derivative structure between 
$\delta^g_{p^-}\bar\phi^a$ and $\delta^g_{q_-}\bar\phi^a$:

\be
\delta^g_{p^-}\bar\phi^a=-\,i\,g\,f^{abc}\frac{1}{\partial^+} 
(\partial\bar\phi^b\partial^+\bar\phi^c)\,; \qquad
\delta^g_{q_-}\bar\phi^a=-g\,f^{abc}\frac{1}{\partial^+}( 
q_+\bar\phi^b\partial^+\bar\phi^c)\ ,
\ee
implies that

\bea
\begin{split}
\int\frac{1}{\partial^+}\phi^a\delta^g_{p^-}\delta^g_{q_-}\bar\phi^a \Bigl|_{\mbox{$\partial$}}
&=~~2\int\,\frac{1}{\partial^+}\delta^g_{p^-}\phi^a\delta^g_{q_-}\bar\phi^a\Bigl|_{\mbox{$\partial$}}\ ,\\
&=-2\int\,\frac{1}{\partial^+}\delta^g_{q_-}\phi^a\delta^g_{p^-}\bar\phi^a\Bigl|_{\mbox{$\partial$}}\ ,
\end{split}
\eea
and leads to

\be
\label{proof2}
\left.\delta^g_{q_-}\,{ H}^g_{\;\partial}~=~-\,2\,i\int 
\frac{1}{\parp}\phi^a\,\delta^g_{p^-}\,\delta^g_{q_-}\,{\bar \phi}^a\,\right|_{\mbox{$\partial$-part}}\ .
\ee
Equating Eqs.~(\ref{proof1}) and (\ref{proof2}) gives

\be
\left.\delta^g_{q_-}\,{ H}^g_{\;\partial}~=~\,i\int\frac{1}{\parp}\phi^a\,[\,\delta^g_{q_-},\,\delta^g_{p^-}\,]\,{\bar \phi}^a\,\right|_{\mbox{$\partial$-part}}\ .
\ee
The next step is to compute

\bea
\label{jaco}
&&\left. i\,[\,\delta^g_{q_-},\,\delta^g_{p^-}\,]\,\bar\phi^a\,\,\right| 
_{\mbox{$\partial$-part}}~= \\
&&=g^2f^{abc}f^{bde}\frac{1}{\partial^+} \biggl\{\frac{\partial}{\partial^+} 
(q_+\bar\phi^d\partial^+\bar\phi^e)
\partial^+\bar\phi^c \biggr\}
+g^2 f^{abc}f^{cde}  \frac{1}{\partial^+}  \biggl\{ \partial\bar\phi^b 
(q_+\bar\phi^d \partial^+\bar\phi^e)\biggr\}\ ,  \nonumber \\
&&\,\,\,-g^2f^{abc}f^{bde} \frac{1}{\partial^+} \biggl\{ \frac{ 
q_+}{\partial^+}\frac{}{}
(\partial\bar\phi^d \partial^+\bar\phi^e) \partial^+\bar\phi^c  \biggr\}
-g^2f^{abc}f^{cde}\frac{1}{\partial^+} \biggl\{ q_+\bar\phi^b 
(\partial\bar\phi^d\partial^+\bar\phi^e)\biggr\} \ .\nonumber
\eea
Expanding the first and third terms yields

\be
g^2f^{abc}f^{bde}\frac{1}{\partial^+}\biggl\{\partial\bar\phi^e 
q_+\bar\phi^d\partial^+\bar\phi^c\biggr\}\ .
\ee
Switching the $b$ and $d$ indices in the last term of Eq.~(\ref{jaco}), allows us to combine it with the second term as

\be
g^2(f^{abc}f^{cde}-f^{adc}f^{cbe})\frac{1}{\partial^+} \biggl\{ \partial\bar\phi^b q_+\bar\phi^d 
\partial^+\bar\phi^e\biggr\} \ .
\ee
The antisymmetry of the structure constants $f^{abc}$, and the {\it Jacobi identity} 

\be
f^{abc}f^{cde}+f^{adc}f^{ceb}+f^{aec}f^{cbd}=0 \ ,
\ee
are necessary to further simplify this equation to 

\be
g^2f^{ace}f^{cbd}\frac{1}{\partial^+}\biggl\{\partial\bar\phi^b 
q_+\bar\phi^d\partial^+\bar\phi^e\biggr\}\ .
\ee
Eq.\,(\ref{jaco}) is then reduced to two terms

\bea
&&g^2f^{abc}f^{bde}\frac{1}{\partial^+}\biggl\{\partial\bar\phi^e 
q_+\bar\phi^d\partial^+\bar\phi^c\biggr\}
+g^2f^{ace}f^{cbd}\frac{1}{\partial^+}\biggl\{\partial\bar\phi^b 
q_+\bar\phi^d\partial^+\bar\phi^e\biggr\}\nonumber\\
&&=g^2\left(f^{ace}f^{cdb}+f^{ace}f^{cbd}\right)\frac{1}{\partial^+} 
\biggl\{\partial\bar\phi^b
q_+\bar\phi^d\partial\bar\phi^e\biggr\}\nonumber\\
&&=0 \ .
\eea
We have therefore shown that 

\be
\delta^g_{q_-}\,\,{ H}^g_{\;\partial}\,=\,0\ .
\ee


\subsection*{Useful identities}
We present here, a series of useful identities which have been used at various stages of this paper.

\vskip 0.3cm

\noindent\underline{\bf Identity 1}\\
For any function $X(\phi)$ of chiral superfields and its conjugate $\bar X(\bar \phi)$,

\bea
\int \bar X \frac{1}{\partial^+} X = \frac{i}{4\sqrt2}\int 
\frac{d^m}{\partial^+}\bar X \frac{\bar d_m}{\partial^+} X \ .
\eea
Proving this identity is rather simple: put $\frac{\partial^+}{\partial^+}$ on the $\bar X$ and rewrite the $\partial^+$ in the numerator as $\frac{i}{4\sqrt2}\{\,\bar d_m,\,d^m\}$. Then integration by parts with respect to $\bar d_m$ yields the identity. 

\vskip 0.3cm
\noindent\underline{\bf Identity 2}\\
\bea
\label{id2}
f^{abc}\int\frac{1}{\partial^+{}^2}\bar\phi^a\phi^b X^c=0\ .
\eea
Use the inside-out relation on $\phi^b$ followed by integration by parts with respect to $(d)^4$. Then swap indices $a$ and $b$ and use the antisymmetry of $f^{abc}$ to obtain Eq. (\ref{id2}). As a corollary, 
\bea
f^{abc}\,\int\,\frac{1}{\partial^+{}^2}\bar\phi^a\partial^+\phi^bX^c \,=\, 
-\,f^{abc}\,\int\,\frac{1}{\partial^+}\bar\phi^a\phi^bX^c\ .
\eea
\vskip 0.3cm

\noindent{\underline{\bf 3-pt function identity}}

\bea
f^{abc} \int \phi^a \frac{\partial}{\partial^+{}^3}{\bar q}_+{}_m (d^m 
{\bar \phi}^b \partial^+{\bar \phi}^c)
=\frac{4i\sqrt{2}}{3}f^{abc}\int\frac{1}{\partial^+}\phi^a{\bar 
\phi}^b\partial{\bar \phi}^c\label{identity}\ .
\eea
This identity is essential to show that the Hamiltonian is a quadratic form. It can be verified by using the explicit forms of $\bar q_+$ and $d$, 
partial integrations, and the inside-out relations. The procedure is as follows (the integral is omitted). 

\begin{itemize}

\item The first step is to replace  $\bar q_+$ and $d$ by $\bar\theta$ and 
$\frac{\partial}{\partial\bar\theta}$, respectively:

\be
f^{abc}\phi^a \frac{\partial}{\partial^+{}^3}{\bar q}_+{}_m (d^m {\bar 
\phi}^b \partial^+{\bar \phi}^c)\\
= i\sqrt2 
f^{abc}\phi^a\frac{\partial}{\partial^+{}^2}\left(\bar\theta\frac{\partial}{\partial\bar\theta}
\bar\phi^b\partial^+\bar\phi^c\right)\ .\label{start}
\ee

\item Perform the partial integration with respect to $\partial^+$ and $\frac{\partial}{\partial\bar\theta}$:

\be
\begin{split}
&4i\sqrt2 
f^{abc}\frac{1}{\partial^+{}^2}\phi^a\partial\bar\phi^b\partial^+\bar\phi^c
-i\sqrt2f^{abc}\frac{1}{\partial^+{}^2}\bar\theta\frac{\partial}{\partial\bar\theta}\phi^a\partial\bar\phi^b\partial^+\bar\phi^c\\
&-i\sqrt2f^{abc}\bar\theta\frac{\partial}{\partial\bar\theta}\bar\phi^b\frac{1}{\partial^+}\phi^a\partial\bar\phi^c
\ ,
\end{split}
\ee
and call the resulting terms $I$, $II$, and $III$, respectively.

\item Work with term $I$: integration by parts with respect to $\parp$ yields
\be
\label{I}
4i\sqrt2 f^{abc}\frac{1}{\partial^+}\phi^a\bar\phi^b\partial\bar\phi^c\,-\,4i\sqrt2 
\frac{1}{\partial^+{}^2}\phi^a\bar\phi^c\partial^+\partial\bar\phi^b \ .
\ee
The second term vanishes thanks to Eq.~(\ref{id2}). Term $I$ is then, 
\bea
I=4i\sqrt2 f^{abc}\frac{1}{\partial^+}\phi^a\bar\phi^b\partial\bar\phi^c\ .
\eea
\item Impose the chiral conditions on term $II$:
\bea
d\phi=0~&\Rightarrow&~-\frac{\partial}{\partial\bar\theta}\phi=\frac{i}{\sqrt2}\theta\partial^+\phi\ ,\\
\bar
d\bar\phi=0~&\Rightarrow&~-\frac{\partial}{\partial\theta}\bar\phi=\frac{i}{\sqrt2}\bar\theta\partial^+\bar\phi\ ,
\eea
These imply that
\be
II=-i\sqrt2f^{abc}\theta\frac{\partial}{\partial\theta}\bar\phi^b\frac{1}{\partial^+}\phi^a\partial\bar\phi^b\partial^+\bar\phi^c
\ .
\ee
Combining $II$ and $III$ gives us
\be
-i\sqrt2f^{abc}(\theta\frac{\partial}{\partial\theta}+{\bar\theta}\frac{\partial}{\partial\bar\theta})\bar\phi^b\frac{1}{\partial^+}\phi^a\partial\bar\phi^c\ .
\ee

\item Use the inside-out relations on ${\bar \phi}^c$ followed by the commutation relation
\be
[ \,\bar d_m, \, 
\theta\frac{\partial}{\partial\theta}+\bar\theta\frac{\partial}{\partial\bar\theta}\,]\,=\,\bar 
q_+{}_m \ ,
\ee
to obtain
\be
II+III=-2i\sqrt2
f^{abc}\phi^a\frac{\partial}{\partial^+{}^2}\left(\bar\theta\frac{\partial}{\partial\bar\theta}\bar\phi^b\partial^+\bar\phi^c\right)
\ ,
\ee
\item Equating Eq.~(\ref{start}) and the sum of terms $I$, $II$ and $III$ proves the 3-pt function identity.
\end{itemize}

\vskip 0.3cm

\noindent{\underline{\bf 4-pt function identity}}\\

\bea
\begin{split}
&\int g^2f^{abc}_{}\,f^{ade}_{}\Big\{\,\frac{1}{\partial^+_{}}(\phi^b\,\partial^+\phi^c)\frac{1}{\partial^+_{}}\,(\bar
\phi^d_{}\,\partial^+_{}\,\bar\phi^e)+\frac{1}{2}\,\phi^b_{}\bar\phi^c\,\phi^d_{}\,\bar\phi^e\Big\} \\
&=\int
\frac{i}{\sqrt{2}} g^2f^{abc}f^{ade}\frac{1}{\partial^+}({\bar d}_m\phi^b\partial^+\phi^c)\frac{1}{\partial^+{}^2}
(d^m{\bar \phi}^d\partial^+{\bar \phi}^e)\ .
\label{fourpt}\end{split}
\eea
We have not yet succeeded in producing an analytic proof of this identity, but we have a proof in terms of the component fields. We checked explicitly that the four-scalar interaction,  after integration over the Grassmann variables, 

\bea\label{4point}
\int-\frac{1}{6}\,g^2\,f^{abc}f^{ade}\frac{1}{\partial^+}\left(\bar C_{pn}{}^{b}\partial^+C^{tn}{}^{c}\right)\frac{1}{\partial^+}\left(C^{pm}{}^{d}\parp\bar C_{tm}{}^{e}\right)\ ,
\eea
is fully reproduced by both sides of Eq. (\ref{fourpt}). Recall that $a,b,c,d,e$ are gauge indices. 

For calculational convenience, we specialize to $SU(2)$. Then this expression splits into three (one for each value of the summed over gauge index $a$), each one containing different fields. We set   $C_{mn}^1\equiv D^{}_{mn}$, and  $C_{mn}^2\equiv E^{}_{mn}$ , and track down terms that involve specific bi-spinor indices such as $\bar D_{12}\,\bar D_{12}$ terms. Then Eq.~(\ref{4point}) becomes

\be
\int \frac{2}{3}\,g^2\,\frac{1}{\partial^+}(\bar D_{12} \partial^+ E^{12})\frac{1}{\partial^+}(E^{12}\partial^+\bar D_{12} ) \ .
\ee
This term exactly matches the $\bar D_{12}\, \bar D_{12}$ terms which come from the components

\bea
&&\int-\frac{1}{8}g^2\,f^{abc}f^{ade}\,\frac{1}{\partial^+}(\partial^+ \bar C_{mn}{}^b C^{mn}{}^c) \frac{1}{\partial^+}(\partial^+ \bar C_{pq}{}^d C^{pq}{}^e) \nonumber\\
&&\quad-\frac{1}{16}g^2\,f^{abc}f^{ade} \,C^{mn}{}^b C^{pq}{}^c \bar C_{mn}{}^d \bar C_{pq}{}^e
\ .
\eea
The algebra is lengthy and not particularly revealing, although the result is non-trivial.
Once we have shown it holds for a particular component, we can use the {\em kinematical} supersymmetry variations to produce the other terms, and thus show the veracity of this claim for all components.

\end{document}